\documentstyle[amssymb,aps,preprint]{revtex}

\begin{document}
\author{Remo Garattini}
\address{M\'{e}canique et Gravitation, Universit\'{e} de Mons-Hainaut,\\
Facult\'e des Sciences, 15 Avenue Maistriau, \\
B-7000 Mons, Belgium \\
and\\
Facolt\`a di Ingegneria, Universit\`a degli Studi di Bergamo,\\
Viale Marconi, 5, 24044 Dalmine (Bergamo) Italy\\
e-mail: Garattini@mi.infn.it}
\title{A Spacetime Foam approach to the cosmological constant and entropy }
\date{\today}
\maketitle

\begin{abstract}
A simple model of spacetime foam, made by $N$ wormholes in a semiclassical
approximation, is taken under examination. The Casimir-like energy of the
quantum fluctuation of such a model and its probability of being realized
are computed. Implications on the Bekenstein-Hawking entropy and the
cosmological constant are considered.
\end{abstract}

\section{Introduction}

The problem of merging General Relativity with Quantum Field Theory is known
as Quantum Gravity. Many efforts to give meaning to a quantum theory of
gravity have been done. Unfortunately until now, such a theory does not
exist, even if other approaches like string theory and the still unknown
M-theory seem to receive a wide agreement in this direction. Nevertheless a
direct investigation of Quantum Gravity shows a lot of interesting aspects.
One of these is that at the Planck scale, spacetime could be subjected to
topology and metric fluctuations\cite{Wheeler}. Such a fluctuating spacetime
is known under the name of ``{\it spacetime foam}'' which can be taken as a
model for the quantum gravitational vacuum\footnote{%
It is interesting to note that there are also indications on how a foamy
spacetime can be tested experimentally\cite{GAC}.}. At this scale of lengths
(or energies) quantum processes like black hole pair creation could become
relevant. To establish if a foamy spacetime could be considered as a
candidate for a Quantum Gravitational vacuum, we can examine the structure
of the effective potential for such a spacetime. It has been shown that flat
space is the classical minimum of the energy for General Relativity\cite{Yau}%
. However there are indications that flat space is not the true ground state
when a temperature is introduced, at least for the Schwarzschild space in
absence of matter fields\cite{GPY}. It is also argued that when gravity is
coupled to $N$ conformally invariant scalar fields the evidence that the
ground-state expectation value of the metric is flat space is false\cite%
{HartleHorowitz}. Moreover it is also believed that in a foamy spacetime,
general relativity can be renormalized when a density of virtual black holes
is taken under consideration coupled to $N$ fermion fields in a $1/N$
expansion\cite{CraneSmolin}. With these examples at hand, we have
investigated the possibility of having a ground state different from flat
space even at zero temperature and what we have discovered is that there
exists an imaginary contribution to the effective potential (more precisely
effective energy) at one loop in a Schwarzschild background, that it means
that flat space is unstable\cite{Remo}. What is the physical interpretation
associated to this instability. We can begin by observing that the
``simplest'' quantum process approximating a foamy spacetime, in absence of
matter fields, could be a black hole pair creation of the neutral type. One
possibility of describing such a process is represented by the
Schwarzschild-de Sitter metric (SdS) which asymptotically approaches the de
Sitter metric. Its degenerate or extreme version is best known as the Nariai
metric\cite{Nariai}. Here we have an external background, the cosmological
constant $\Lambda $, which gives a nonzero probability of having a neutral
black hole pair produced with its components accelerating away from each
other. Nevertheless this process is believed to be highly suppressed, at
least for $\Lambda \gg 1$ in Planck's units\cite{Bousso-Hawking}. In any
case, metrics with a cosmological constant have different boundary
conditions compared to flat space. The Schwarzschild metric is the only case
available. Here the whole spacetime can be regarded as a black
hole-anti-black hole pair formed up by a black hole with positive mass $M$
in the coordinate system of the observer and an {\it anti black-hole} with
negative mass $-M$ in the system where the observer is not present. In this
way we have an energy preserving mechanism, because flat space has {\it zero
energy} and the pair has zero energy too. However, in this case we have not
a cosmological {\it force} producing the pair: we have only pure
gravitational fluctuations. The black hole-anti-black hole pair has also a
relevant pictorial interpretation: the black hole with positive mass $M$ and
the {\it anti black-hole} with negative mass $-M$ can be considered the
components of a virtual dipole with zero total energy created by a large
quantum gravitational fluctuation\cite{Modanese}. Note that this is the only
physical process compatible with the energy conservation. The importance of
having the same energy behaviour ({\it asymptotic}) is related to the
possibility of having a spontaneous transition from one spacetime to another
one with the same boundary condition \cite{Witten}. This transition is a
decay from the false vacuum to the true one\cite{Coleman,Mazur}. However, if
we take account of a pair of neutral black holes living in different
universes, there is no decay and more important no temperature is involved
to change from flat to curved space. To see if this process is realizable we
need to compute quantum corrections to the energy stored in the boundaries.
These quantum corrections are pure gravitational vacuum excitations which
can be measured by the Casimir energy, formally defined as
\begin{equation}
E_{Casimir}\left[ \partial {\cal M}\right] =E_{0}\left[ \partial {\cal M}%
\right] -E_{0}\left[ 0\right] ,  \label{i0}
\end{equation}
where $E_{0}$ is the zero-point energy and $\partial {\cal M}$ is a
boundary. For zero temperature, the idea underlying the Casimir effect is to
compare vacuum energies in two physical distinct configurations. For
gravitons embedded in flat space, the one-loop contribution to the zero
point energy (ZPE) is
\begin{equation}
2\cdot \frac{1}{2}\int \frac{d^{3}k}{\left( 2\pi \right) ^{3}}\sqrt{k^{2}}.
\end{equation}
This term is UV\ divergent, and its effect is equivalent to inducing a
divergent ``{\it cosmological constant''}\cite{BES,BES1}
\begin{equation}
\Lambda _{ZPE}=8\pi G\rho _{ZPE}=4\pi G\int \frac{d^{3}k}{\left( 2\pi
\right) ^{3}}\sqrt{k^{2}}.  \label{i1}
\end{equation}
However, it is likely that a natural cutoff of the order of the inverse
Planck length comes into play. If this is the case we expect
\begin{equation}
\Lambda _{ZPE}=c\frac{8\pi }{l_{P}^{2}},
\end{equation}
where $c$ is a dimensionless constant. There are some observational data
coming from the Friedmann-Robertson-Walker cosmology constraining the
cosmological constant\cite{MVisser}. An estimate of these ones is
\begin{equation}
\begin{array}{c}
\Lambda \lesssim 10^{-122}l_{P}^{-2} \\
c\lesssim 10^{-123}%
\end{array}
.  \label{i2}
\end{equation}
What is the relation between $\Lambda _{ZPE}$ and the neutral black hole
pair production? Since the pair produced is mediated by a three-dimensional
wormhole with its own ZPE showing a Casimir-like energy and an imaginary
part, if we enlarge the number of wormholes (or equivalently the pair
creation number) to a certain value $N$. In section \ref{p2}, we will prove
that the imaginary part disappears and the system from unstable turns to be
stable. In Ref.\cite{Remo1} a spacetime foam model made by $N$ {\it coherent}
wormholes has been proposed. In that paper, we have computed the energy
density of gravitational fluctuations reproducing the same behavior
conjectured by Wheeler during the sixties on dimensional grounds\cite%
{Wheeler}. As an application of the model proposed in Ref.\cite{Remo1}, in
Ref.\cite{Remo2} we have computed the spectrum of a generic area and as a
particular case we have considered the de Sitter geometry. The result is a
{\it quantization} process whose quanta can be identified with wormholes of
Planckian size. In this paper, we would like to continue the investigation
of such a model of {\it spacetime foam}, by looking at the problem of the
cosmological constant, from the Casimir-like energy point of view, and as a
consequence of the {\it quantization} process of Ref.\cite{Remo2}, we will
give indications about the black hole mass quantization. The rest of the
paper is structured as follows, in section \ref{p2} and \ref{p3} we briefly
recall the results reported in Refs.\cite{Remo1,Remo2}, by looking at the
mass quantization process. In section \ref{p4} we approach the cosmological
constant problem. We summarize and conclude in section \ref{p5}. Units in
which $\hbar =c=k=1$ are used throughout the paper.

\section{Spacetime foam: the model}

\label{p2}In the one-wormhole approximation we have used an eternal black
hole, to describe a complete manifold ${\cal M}$, composed of two wedges $%
{\cal M}_{+}$ and ${\cal M}_{-}$ located in the right and left sectors of a
Kruskal diagram. The spatial slices $\Sigma $ represent Einstein-Rosen
bridges with wormhole topology $S^{2}\times R^{1}$. Also the hypersurface $%
\Sigma $ is divided in two parts $\Sigma _{+}$ and $\Sigma _{-}$ by a
bifurcation two-surface $S_{0}$. The line element we shall consider is
\begin{equation}
ds^{2}=-N^{2}\left( r\right) dt^{2}+\frac{dr^{2}}{1-\frac{2MG}{r}}%
+r^{2}\left( d\theta ^{2}+\sin ^{2}\theta d\phi ^{2}\right)   \label{p21}
\end{equation}%
or by defining the proper distance from the throat $dy=\pm dr/\sqrt{1-\frac{%
2MG}{r}}$, metric $\left( \ref{p21}\right) $ becomes
\begin{equation}
ds^{2}=-N^{2}dt^{2}+g_{yy}dy^{2}+r^{2}\left( y\right) d\Omega ^{2}.
\end{equation}%
$N$, $g_{yy}$, and $r$ are functions of the radial coordinate $y$
continuously defined on ${\cal M}$. The throat of the bridge is at $r=2MG$
or $y=0$. We choose $y$ to be positive in $\Sigma _{+}$ and negative in $%
\Sigma _{-}$. If we make the identification $N^{2}=1-2MG/r$, the line
element $\left( \ref{p21}\right) $ reduces to the Schwarzschild metric
written in another form. The boundaries $^{2}S_{+}$ and $^{2}S_{-}$ are
located at coordinate values $y=y_{+}$ and $y=y_{-}$ respectively. The
physical Hamiltonian defined on $\Sigma $ is
\begin{equation}
H_{P}=H-H_{0}=\frac{1}{16\pi l_{p}^{2}}\int_{\Sigma }d^{3}x\left( N{\cal H}%
+N_{i}{\cal H}^{i}\right) +H_{S_{+}}-H_{S_{-}}
\end{equation}%
where $l_{p}^{2}=G$ and $H_{S_{+}}-H_{S_{-}}$ represents the boundary
hamiltonian defined by
\begin{equation}
+\text{ }\frac{1}{8\pi l_{p}^{2}}\int_{S_{+}}d^{2}xN\sqrt{\sigma }\left(
k-k^{0}\right) -\frac{1}{8\pi l_{p}^{2}}\int_{S_{-}}d^{2}xN\sqrt{\sigma }%
\left( k-k^{0}\right) ,
\end{equation}%
where $\sigma $ is the two-dimensional determinant coming from the induced
metric $\sigma _{ab}$ on the boundaries $S_{\pm }$. The volume term contains
two constraints
\begin{equation}
\left\{
\begin{array}{l}
{\cal H}=G_{ijkl}\pi ^{ij}\pi ^{kl}\left( \frac{l_{p}^{2}}{\sqrt{g}}\right)
-\left( \frac{\sqrt{g}}{l_{p}^{2}}\right) R^{\left( 3\right) }=0 \\
{\cal H}^{i}=-2\pi _{|j}^{ij}=0%
\end{array}%
\right. ,
\end{equation}%
both satisfied by the Schwarzschild and Flat metric on-shell, respectively.
The supermetric is $G_{ijkl}=\frac{1}{2}\left(
g_{ik}g_{jl}+g_{il}g_{jk}-g_{ij}g_{kl}\right) $ and $R^{\left( 3\right) }$
denotes the scalar curvature of the surface $\Sigma $. In particular on the
boundary we shall use the quasilocal energy definition. Quasilocal energy $%
E_{ql}$ is defined as the value of the Hamiltonian that generates unit time
translations orthogonal to the two-dimensional boundaries, i.e.\cite%
{BrownYork,FroMar,HawHor,Martinez}
\[
E_{ql}=E_{+}-E_{-},
\]%
\[
E_{+}=\frac{1}{8\pi l_{p}^{2}}\int_{S_{+}}d^{2}x\sqrt{\sigma }\left(
k-k^{0}\right)
\]%
\begin{equation}
E_{-}=-\frac{1}{8\pi l_{p}^{2}}\int_{S_{-}}d^{2}x\sqrt{\sigma }\left(
k-k^{0}\right) .  \label{p22}
\end{equation}%
where $\left| N\right| =1$ at both $S_{+}$ and $S_{-}$. $E_{ql}$ is the
quasilocal energy of a spacelike hypersurface $\Sigma =\Sigma _{+}\cup
\Sigma _{-}$ bounded by two boundaries $^{3}S_{+}$ and $^{3}S_{-}$ located
in the two disconnected regions $M_{+}$ and $M_{-}$ respectively. We have
included the subtraction terms $k^{0}$ for the energy. $k^{0}$ represents
the trace of the extrinsic curvature corresponding to embedding in the
two-dimensional boundaries $^{2}S_{+}$ and $^{2}S_{-}$ in three-dimensional
Euclidean space. By using the expression of the trace
\begin{equation}
k=-\frac{1}{\sqrt{h}}\left( \sqrt{h}n^{\mu }\right) _{,\mu },
\end{equation}%
with the normal to the boundaries defined continuously along $\Sigma $ as $%
n^{\mu }=\left( h^{yy}\right) ^{\frac{1}{2}}\delta _{y}^{\mu }$, the value
of $k$ depends on the function $r,_{y}$, where we have assumed that the
function $r,_{y}$ is positive for $S_{+}$ and negative for $S_{-}$. We
obtain at either boundary that
\begin{equation}
k=\frac{-2r,_{y}}{r}.
\end{equation}%
The trace associated with the subtraction term is taken to be $k^{0}=-2/r$
for $B_{+}$ and $k^{0}=2/r$ for $B_{-}$. Thus the quasilocal energy with
subtraction terms included is
\begin{equation}
E_{ql}=\left( E_{+}-E_{-}\right) =l_{p}^{-2}\left[ \left( r\left[ 1-\left|
r,_{y}\right| \right] \right) _{y=y_{+}}-\left( r\left[ 1-\left|
r,_{y}\right| \right] \right) _{y=y_{-}}\right] .
\end{equation}%
It is easy to see that $E_{+}$ and $E_{-}$ tend individually to the ${\cal %
ADM}$ mass $M$ when the boundaries $^{3}B_{+}$ and $^{3}B_{-}$ tend
respectively to right and left spatial infinity. It should be noted that the
total energy is zero for boundary conditions symmetric with respect to the
bifurcation surface, i.e.,
\begin{equation}
E=E_{+}-E_{-}=M+\left( -M\right) =0,  \label{p24}
\end{equation}%
where the asymptotic contribution has been considered. We can recognize that
the expression which defines quasilocal energy is formally of the Casimir
type. Indeed, the subtraction procedure present in Eq.$\left( \ref{p22}%
\right) $ describes an energy difference between two distinct situations
with the same boundary conditions. The same behaviour appears in the entropy
calculation for the physical system under examination. Indeed\cite{Martinez}
\begin{equation}
S_{tot}=S_{+}-S_{-}=\frac{A^{+}}{4l_{p}^{2}}-\frac{A^{-}}{4l_{p}^{2}}\simeq
\frac{A_{H}}{4l_{p}^{2}}-\frac{A_{H}}{4l_{p}^{2}}=0,  \label{p25}
\end{equation}%
where $A^{+}$ and $A^{-}$ have the same meaning as $E_{+}$ and $E_{-}$. Note
that for both entropy and energy this result is obtained at the tree level.
In particular, the quasilocal energy can be interpreted as the tree level
approximation of the Casimir energy. We can also see Eqs.$\left( \ref{p24}%
\right) $ and $\left( \ref{p25}\right) $ from a different point of view. In
fact these equations say that flat space can be thought of as a composition
of two pieces: the former, with positive energy, in the region $\Sigma _{+}$
and the latter, with negative energy, in the region $\Sigma _{-}$, where the
positive and negative concern the bifurcation surface (hole) which is formed
due to a topology change of the manifold. The most appropriate mechanism to
explain this splitting seems to be a neutral black hole pair creation. To
this purpose we begin by considering perturbations at $\Sigma $ of the type
\begin{equation}
g_{ij}^{S,F}=\bar{g}_{ij}^{S,F}+h_{ij},
\end{equation}%
where $\bar{g}_{ij}^{S}$ is the spatial part of the Schwarzschild background
and $\bar{g}_{ij}^{F}$ is the spatial part of the Flat background. In this
framework we compute the effective energy defined by
\begin{equation}
E=\min\limits_{\left\{ \Psi \right\} }\frac{\left\langle \Psi \left|
H\right| \Psi \right\rangle }{\left\langle \Psi |\Psi \right\rangle }
\end{equation}%
with the help of the following rules
\begin{equation}
\begin{array}{c}
\int \left[ {\cal D}h_{ij}\right] h_{ij}\left( \overrightarrow{x}\right)
\mid \Psi \left\{ h_{ij}+\bar{g}_{ij}\right\} \mid ^{2}=0, \\
\\
\frac{\int \left[ {\cal D}h_{ij}\right] \int d^{3}xh_{ij}\left(
\overrightarrow{x}\right) h_{kl}\left( \overrightarrow{y}\right) \mid \Psi
\left\{ h_{ij}+\bar{g}_{ij}\right\} \mid ^{2}}{\int \left[ {\cal D}h_{ij}%
\right] \mid \Psi \left\{ h_{ij}+\bar{g}_{ij}\right\} \mid ^{2}}%
=K_{ijkl}\left( \overrightarrow{x},\overrightarrow{y}\right) .%
\end{array}%
\end{equation}%
In particular
\begin{equation}
E\left( M\right) =\frac{\left\langle \Psi \left| H^{Schw.}\right| \Psi
\right\rangle }{\left\langle \Psi |\Psi \right\rangle }+\frac{\left\langle
\Psi \left| H_{boundary}^{Schw.}\right| \Psi \right\rangle }{\left\langle
\Psi |\Psi \right\rangle }
\end{equation}%
and
\begin{equation}
E\left( 0\right) =\frac{\left\langle \Psi \left| H^{Flat}\right| \Psi
\right\rangle }{\left\langle \Psi |\Psi \right\rangle }+\frac{\left\langle
\Psi \left| H_{boundary}^{Flat}\right| \Psi \right\rangle }{\left\langle
\Psi |\Psi \right\rangle }
\end{equation}%
so that the physical Hamiltonian is given by the difference (Casimir energy)
$\Delta E\left( M\right) $%
\begin{equation}
=E\left( M\right) -E\left( 0\right) =\frac{\left\langle \Psi \left|
H^{Schw.}-H^{Flat}\right| \Psi \right\rangle }{\left\langle \Psi |\Psi
\right\rangle }+\frac{\left\langle \Psi \left| H_{ql}\right| \Psi
\right\rangle }{\left\langle \Psi |\Psi \right\rangle }.  \label{p26}
\end{equation}%
The quantity $\Delta E\left( M\right) $ is computed by means of a
variational approach, where the WKB functionals are substituted with trial
wave functionals. By restricting our attention to the graviton sector of the
Hamiltonian approximated to second order, hereafter referred as $H_{|2}$, we
define
\[
E_{|2}=\frac{\left\langle \Psi ^{\perp }\left| H_{|2}\right| \Psi ^{\perp
}\right\rangle }{\left\langle \Psi ^{\perp }|\Psi ^{\perp }\right\rangle },
\]%
where
\[
\Psi ^{\perp }=\Psi \left[ h_{ij}^{\perp }\right] ={\cal N}\exp \left\{ -%
\frac{1}{4l_{p}^{2}}\left[ \left\langle \left( g-\bar{g}\right) K^{-1}\left(
g-\bar{g}\right) \right\rangle _{x,y}^{\perp }\right] \right\} .
\]%
After having functionally integrated $H_{|2}$, we get
\begin{equation}
H_{|2}=\frac{1}{4l_{p}^{2}}\int_{\Sigma }d^{3}x\sqrt{g}G^{ijkl}\left[
K^{-1\bot }\left( x,x\right) _{ijkl}+\left( \triangle _{2}\right)
_{j}^{a}K^{\bot }\left( x,x\right) _{iakl}\right]
\end{equation}%
The propagator $K^{\bot }\left( x,x\right) _{iakl}$ comes from a functional
integration and it can be represented as
\begin{equation}
K^{\bot }\left( \overrightarrow{x},\overrightarrow{y}\right)
_{iakl}:=\sum_{N}\frac{h_{ia}^{\bot }\left( \overrightarrow{x}\right)
h_{kl}^{\bot }\left( \overrightarrow{y}\right) }{2\lambda _{N}\left(
p\right) },
\end{equation}%
where $h_{ia}^{\bot }\left( \overrightarrow{x}\right) $ are the
eigenfunctions of
\begin{equation}
\left( \triangle _{2}\right) _{j}^{a}:=-\triangle \delta _{j}^{a}+2R_{j}^{a}.
\end{equation}%
This is the Lichnerowicz operator projected on $\Sigma $ acting on traceless
transverse quantum fluctuations and $\lambda _{N}\left( p\right) $ are
infinite variational parameters. $\triangle $ is the curved Laplacian
(Laplace-Beltrami operator) on a Schwarzschild background and $R_{j\text{ }%
}^{a}$ is the mixed Ricci tensor whose components are:
\begin{equation}
R_{j}^{a}=diag\left\{ \frac{-2MG}{r^{3}},\frac{MG}{r^{3}},\frac{MG}{r^{3}}%
\right\} .
\end{equation}%
The minimization with respect to $\lambda $ and the introduction of a high
energy cutoff $\Lambda $ give to the Eq. $\left( \ref{p26}\right) $ the
following form
\begin{equation}
\Delta E\left( M\right) \sim -\frac{V}{32\pi ^{2}}\left( \frac{3MG}{r_{0}^{3}%
}\right) ^{2}\ln \left( \frac{r_{0}^{3}\Lambda ^{2}}{3MG}\right) ,
\label{p27}
\end{equation}%
where $V$ is the volume of the system and $r_{0}$ is related to the minimum
radius compatible with the wormhole throat. We know that classically, the
minimum radius is achieved when $r_{0}=2MG$. However, it is likely that
quantum processes come into play at short distances, where the wormhole
throat is defined, introducing a {\it quantum} radius $r_{0}>2MG$.
Nevertheless, since we are interested to probe the energy contribution near
the Planck scale we can fix the value of $r_{0}=l_{p}$ from now on. We now
compute the minimum of $\Delta E\left( M\right) $, after having rescaled the
variable $M$ to a scale variable $x=3MG/\left( r_{0}^{3}\Lambda ^{2}\right)
=3M/\left( l_{p}\Lambda ^{2}\right) $. Thus
\[
\Delta E\left( M\right) \rightarrow \Delta E\left( x,\Lambda \right) =\frac{V%
}{32\pi ^{2}}\Lambda ^{4}x^{2}\ln x
\]%
We obtain two values for $x$: $x_{1}=0$, i.e. flat space and $x_{2}=e^{-%
\frac{1}{2}}$. At the minimum
\begin{equation}
\Delta E\left( x_{2}\right) =-\frac{V}{64\pi ^{2}}\frac{\Lambda ^{4}}{e}.
\end{equation}%
Nevertheless, there exists another part of the spectrum which has to be
considered: the discrete spectrum containing one mode. This gives the energy
an imaginary contribution, namely we have discovered an unstable mode\cite%
{GPY,Remo}. Let us briefly recall, how this appears. The eigenvalue equation
\begin{equation}
\left( \triangle _{2}\right) _{i}^{a}h_{aj}=\alpha h_{ij}
\end{equation}%
can be studied with the Regge-Wheeler method. The perturbations can be
divided in odd and even components. The appearance of the unstable mode is
governed by the gravitational field component $h_{11}^{even}$. Explicitly
\[
-E^{2}H\left( r\right)
\]%
\begin{equation}
=-\left( 1-\frac{2MG}{r}\right) \frac{d^{2}H\left( r\right) }{dr^{2}}+\left(
\frac{2r-3MG}{r^{2}}\right) \frac{dH\left( r\right) }{dr}-\frac{4MG}{r^{3}}%
H\left( r\right) ,  \label{p28}
\end{equation}%
where
\begin{equation}
h_{11}^{even}\left( r,\vartheta ,\phi \right) =\left[ H\left( r\right)
\left( 1-\frac{2m}{r}\right) ^{-1}\right] Y_{00}\left( \vartheta ,\phi
\right)
\end{equation}%
and $E^{2}>0$. Eq.$\left( \ref{p28}\right) $ can be transformed into
\begin{equation}
\mu =\frac{\int\limits_{0}^{\bar{y}}dy\left[ \left( \frac{dh\left( y\right)
}{dy}\right) ^{2}-\frac{3}{2\rho \left( y\right) ^{3}}h^{2}\left( y\right) %
\right] }{\int\limits_{0}^{\bar{y}}dyh^{2}\left( y\right) },
\end{equation}%
where $\mu $ is the eigenvalue, $y$ is the proper distance from the throat
in dimensionless form. If we choose $h\left( \lambda ,y\right) =\exp \left(
-\lambda y\right) $ as a trial function we numerically obtain $\mu =-.701626$%
. In terms of the energy square we have
\begin{equation}
E^{2}=-.\,\allowbreak 175\,41/\left( MG\right) ^{2}
\end{equation}%
to be compared with the value $E^{2}=-.\,\allowbreak 19/\left( MG\right) ^{2}
$ of Ref.\cite{GPY}. Nevertheless, when we compute the eigenvalue as a
function of the distance $y$, we discover that in the limit $\bar{y}%
\rightarrow 0$,
\begin{equation}
\mu \equiv \ \mu \left( \lambda \right) =\lambda ^{2}-\frac{3}{2}+\frac{9}{8}%
\left[ \bar{y}^{2}+\frac{\bar{y}}{2\lambda }\right] .
\end{equation}%
Its minimum is at $\tilde{\lambda}=\left( \frac{9}{32}\bar{y}\right) ^{\frac{%
1}{3}}$ and
\begin{equation}
\mu \left( \tilde{\lambda}\right) =1.\,\allowbreak 287\,8\bar{y}^{\frac{2}{3}%
}+\frac{9}{8}\bar{y}^{2}-\frac{3}{2}.
\end{equation}%
It is evident that there exists a critical radius where $\mu $ turns from
negative to positive. This critical value is located at $\rho
_{c}=1.\,\allowbreak 113\,4$ to be compared with the value $\rho _{c}=1.445$
obtained by B. Allen in \cite{B.Allen}. What is the relation with the large
number of wormholes? As mentioned in Ref.\cite{Remo1}, when the number of
wormholes grows, the space available for every single wormhole has to be
reduced to avoid overlapping of the wave functions. Consider the simple case
of two wormholes covering the hypersurface $\Sigma $, namely $\Sigma =\Sigma
_{1}\cup \Sigma _{2}$, $\Sigma _{1}\cap \Sigma _{2}=\emptyset $. The second
property assures that the boundaries and the support of the wave functional
do not overlap, in agreement with the WKB approximation. $\Sigma _{1}$and $%
\Sigma _{2}$ have topology $S^{2}\times R^{1}$ with boundaries $\partial
\Sigma _{1}^{\pm }$ and $\partial \Sigma _{2}^{\pm }$ with respect to each
bifurcation surface. The total Hamiltonian, in this case, is $%
H_{T}=H_{1}+H_{2}$, i.e. (here we are looking at boundary terms, because in
this discussion they are the only relevant ones)
\[
H_{T}=2H=\frac{1}{8\pi l_{p}^{2}}\left[ 2\int_{S_{+}}d^{2}x\sqrt{\sigma }%
\left( k-k^{0}\right) -2\int_{S_{-}}d^{2}x\sqrt{\sigma }\left(
k-k^{0}\right) \right]
\]%
\[
=\frac{1}{8\pi l_{p}^{2}}\left[ 2\left( r\left[ 1-\left| r,_{y}\right| %
\right] \right) _{y=y_{+}}-2\left( r\left[ 1-\left| r,_{y}\right| \right]
\right) _{y=y_{-}}\right]
\]%
\[
=\frac{1}{8\pi l_{p}^{2}}\left[ 2r_{+}\left( 1-\sqrt{1-\frac{2Ml_{p}^{2}}{%
r_{+}}}\right) -2r_{-}\left( 1-\sqrt{1-\frac{2Ml_{p}^{2}}{r_{-}}}\right) %
\right]
\]%
\[
=\frac{1}{8\pi l_{p}^{2}}\left[ 2r_{+}\left( 1-\sqrt{1-\frac{2Ml_{2_{w}}^{2}%
}{2r_{+}}}\right) -2r_{-}\left( 1-\sqrt{1-\frac{2Ml_{2_{w}}^{2}}{2r_{-}}}%
\right) \right]
\]%
\begin{equation}
=\frac{1}{8\pi l_{p}^{2}}\left[ R_{+}\left( 1-\sqrt{1-\frac{2Ml_{2_{w}}^{2}}{%
R_{+}}}\right) -R_{-}\left( 1-\sqrt{1-\frac{2Ml_{2_{w}}^{2}}{R_{-}}}\right) %
\right] .
\end{equation}%
This means that the total quasilocal energy is the same of a single wormhole
with boundaries satisfying the relation $R_{\pm }=2r_{\pm }$, or in other
words the value of $R_{\pm }$ in presence of two wormholes is divided by
two. Note that $R_{\pm }$ are the boundary values corresponding to the
single wormhole case. This implies that if we put more and more wormholes,
say $N_{w}$, the initial boundary located at $R_{\pm }$ will be reduced and $%
G\rightarrow N_{w}G$\label{page}. This boundary reduction is important,
because it is related to the disappearing of the unstable mode. Let us see
how. If we fix the initial boundary at $R_{\pm }$, then in presence of $N_{w}
$ wormholes, it will be reduced to $R_{\pm }/N_{w}$. This means that
boundary conditions are not fixed at infinity, but at a certain finite
radius and the $ADM$ mass term is substituted by the quasilocal energy
expression under the condition of having symmetry with respect to each
bifurcation surface. The effect on the unstable mode is clear: as $N_{w}$
grows, the boundary radius reduces more and more until it will reach the
critical value $\rho _{c}$ below which no negative mode will appear
corresponding to a critical wormholes number $N_{w_{c}}$. To this purpose,
suppose to consider $N_{w}$ wormholes and assume that there exists a
covering of $\Sigma $ such that $\Sigma =\bigcup\limits_{i=1}^{N_{w}}\Sigma
_{i}$, with $\Sigma _{i}\cap \Sigma _{j}=\emptyset $ when $i\neq j$. Each $%
\Sigma _{i}$ has the topology $S^{2}\times R^{1}$ with boundaries $\partial
\Sigma _{i}^{\pm }$ with respect to each bifurcation surface. On each
surface $\Sigma _{i}$, quasilocal energy gives\cite%
{BrownYork,FroMar,HawHor,Martinez}
\begin{equation}
E_{i\text{ }{\rm ql}}=\frac{1}{8\pi l_{p}^{2}}\int_{S_{i+}}d^{2}x\sqrt{%
\sigma }\left( k-k^{0}\right) -\frac{1}{8\pi l_{p}^{2}}\int_{S_{i-}}d^{2}x%
\sqrt{\sigma }\left( k-k^{0}\right) .
\end{equation}%
Thus if we apply the same procedure of the single case on each wormhole, we
obtain
\begin{equation}
E_{i\text{ }{\rm ql}}=\left( E_{i+}-E_{i-}\right) =l_{p}^{-2}\left( r\left[
1-\left| r,_{y}\right| \right] \right) _{y=y_{i+}}-l_{p}^{-2}\left( r\left[
1-\left| r,_{y}\right| \right] \right) _{y=y_{i-}}.
\end{equation}%
Note that the total quasilocal energy is zero for boundary conditions
symmetric with respect to {\it each} bifurcation surface $S_{0,i}$. If we
assume this kind of symmetry for boundary conditions, it is immediate to
recognize that the vanishing of the boundary term is guaranteed beyond the
semiclassical approximation, because every term on one wedge of the
hypersurface $\Sigma _{i}$ will be compensated by the term on the other
wedge of the same hypersurface $\Sigma _{i}$, giving therefore zero energy
contribution. We are interested in a large number of wormholes, each of them
contributing with a term of the type $E_{i\text{ }{\rm ql}}$. If the
wormholes number is $N_{w}$, we obtain (semiclassically, i.e., without
self-interactions)
\begin{equation}
H_{tot}^{N_{w}}=H^{1}+H^{2}+\ldots +H^{N_{w}}.
\end{equation}%
Thus the total energy for the collection is
\[
E_{|2}^{tot}=N_{w}H_{|2}.
\]%
The same happens for the trial wave functional which is the product of $N_{w}
$ t.w.f.. Thus
\begin{equation}
\Psi _{tot}^{\perp }=\Psi _{1}^{\perp }\otimes \Psi _{2}^{\perp }\otimes
\ldots \ldots \Psi _{N_{w}}^{\perp }={\cal N}\exp N_{w}\left\{ -\frac{1}{%
4l_{p}^{2}}\left[ \left\langle \left( g-\bar{g}\right) K^{-1}\left( g-\bar{g}%
\right) \right\rangle _{x,y}^{\perp }\right] \right\} .
\end{equation}%
Thus for the $N_{w}$ wormholes, one gets%
\begin{equation}
\Delta E_{N_{w}}\left( x,\Lambda \right) \sim N_{w}\frac{V}{32\pi ^{2}}%
\Lambda ^{4}x^{2}\ln x,  \label{p28a}
\end{equation}%
where we have defined the usual scale variable $x=3M/\left( l_{p}\Lambda
^{2}\right) $. Then at one loop the cooperative effects of wormholes behave
as one {\it macroscopic single }field multiplied by $N_{w}^{2}$, but without
the unstable mode. At the minimum, $\bar{x}=e^{-\frac{1}{2}}$%
\begin{equation}
\Delta E\left( \bar{x}\right) =-N_{w}\frac{V}{64\pi ^{2}}\frac{\Lambda ^{4}}{%
e}.  \label{p29}
\end{equation}%
This means that we have obtained a minimum of the effective energy away by
the flat space, indicating that another configuration has to be considered
for the ground state of quantum gravity. Let us examine the implications on
the area quantization, entropy and the cosmological constant.

\section{Area spectrum and Entropy}

\label{p3}Bekenstein made the proposal that a black hole {\it does} have an
entropy proportional to the area of its horizon\cite{J.Bekenstein}
\begin{equation}
S_{bh}=const\times A_{hor}.
\end{equation}
In particular, in natural units one finds that the proportionality constant
is set to $1/4G=1/4l_{p}^{2}$, so that the entropy becomes
\begin{equation}
S=\frac{A}{4G}=\frac{A}{4l_{p}^{2}}.
\end{equation}
Following Bekenstein's proposal on the quantization of the area for
nonextremal black holes we have
\begin{equation}
A_{n}=\alpha l_{p}^{2}\left( n+\eta \right) \text{\qquad }\eta >-1\text{%
\qquad }n=1,2,\ldots  \label{p30a}
\end{equation}
Many attempts to recover the area spectrum have been done, see Refs.\cite%
{Bekenstein,Mukohyama} for a review. Note that the appearance of a discrete
spectrum is not so trivial. Indeed there are other theories, based on
spherically symmetric metrics in a mini-superspace approach, whose mass
spectrum is continuous\cite{Hollmann,Cavaglià}. The area is measured by the
quantity
\begin{equation}
A\left( S_{0}\right) =\int_{S_{0}}d^{2}x\sqrt{\sigma }.
\end{equation}
$\sigma $ is the two-dimensional determinant coming from the induced metric $%
\sigma _{ab}$ on the boundary $S_{0}$. We would like to evaluate the mean
value of the area
\begin{equation}
A\left( S_{0}\right) =\frac{\left\langle \Psi _{F}\left| \hat{A}\right| \Psi
_{F}\right\rangle }{\left\langle \Psi _{F}|\Psi _{F}\right\rangle }=\frac{%
\left\langle \Psi _{F}\left| \widehat{\int_{S_{0}}d^{2}x\sqrt{\sigma }}%
\right| \Psi _{F}\right\rangle }{\left\langle \Psi _{F}|\Psi
_{F}\right\rangle },
\end{equation}
computed on
\begin{equation}
\left| \Psi _{F}\right\rangle =\Psi _{1}^{\perp }\otimes \Psi _{2}^{\perp
}\otimes \ldots \ldots \Psi _{N_{w}}^{\perp }.  \label{p30}
\end{equation}
Since we are working with spherical symmetric wormholes we consider $\sigma
_{ab}=\bar{\sigma}_{ab}+\delta \sigma _{ab}$, where $\bar{\sigma}_{ab}$ is
such that $\int_{S_{0}}d^{2}x\sqrt{\bar{\sigma}}=4\pi \bar{r}^{2}$ and $\bar{%
r}$ is the radius of $S_{0}$. To the lowest level in the expansion of $%
\sigma _{ab}$ we obtain that
\begin{equation}
A\left( S_{0}\right) =\frac{\left\langle \Psi _{F}\left| \hat{A}\right| \Psi
_{F}\right\rangle }{\left\langle \Psi _{F}|\Psi _{F}\right\rangle }=4\pi
\bar{r}^{2}.  \label{p31}
\end{equation}
Suppose to consider the mean value of the area $A$ computed on a given {\it %
macroscopic} fixed radius $R$. On the basis of our foam model, we obtain $%
A=\bigcup\limits_{i=1}^{N}A_{i}$, with $A_{i}\cap A_{j}=\emptyset $ when $%
i\neq j$. Thus
\begin{equation}
A=4\pi R^{2}=\sum\limits_{i=1}^{N}A_{i}=\sum\limits_{i=1}^{N}4\pi \bar{r}%
_{i}^{2}.
\end{equation}
In Refs.\cite{Remo2} we have considered, as a first approximation, the limit
$\bar{r}_{i}\rightarrow l_{p}$ and we have obtained
\begin{equation}
A=NA_{l_{p}}=N4\pi l_{p}^{2}.  \label{p32}
\end{equation}
Nevertheless an improvement of Eq.$\left( \ref{p32}\right) $ is possible if
we introduce a scale variable $x_{i}=\bar{r}_{i}/l_{p}$ which leads to
\begin{equation}
A=4\pi l_{p}^{2}\sum\limits_{i=1}^{N}x_{i}^{2}=4\pi l_{p}^{2}N\overline{x^{2}%
}=4\pi l_{p}^{2}N\alpha .  \label{p33a}
\end{equation}
Thus the number $\alpha $ appearing in Eq.$\left( \ref{p30a}\right) $, here
comes from an averaging process. Note that the $4\pi $ factor is a
consequence of the $S^{2}$ wormhole topology which is an intrinsic feature
of our foam model. Comparison of Eq.$\left( \ref{p33a}\right) $with the
Bekenstein area spectrum\cite{Bekenstein} gives
\begin{equation}
4\pi l_{p}^{2}N\alpha =4l_{p}^{2}N\ln 2.
\end{equation}
This fixes the coefficient $\alpha $ to
\begin{equation}
\frac{\ln 2}{\pi }=\alpha
\end{equation}
and the entropy is
\begin{equation}
S=\frac{A}{4l_{p}^{2}}=\frac{4l_{p}^{2}N\ln 2}{4l_{p}^{2}}=N\ln 2.
\end{equation}
$N$ is such that $N\geq N_{w_{c}}$ and $N_{w_{c}}$ is the critical wormholes
number above which we have the stability of our foam model. On the other
hand if we apply the same reasoning of Refs.\cite{Rovelli,Rovelli1}, applied
to the quantity
\begin{equation}
\frac{A}{4\pi l_{p}^{2}}=\sum\limits_{i=1}^{N}x_{i}^{2},
\end{equation}
produces an extra-factor of the form $\ln 2/\pi $, when compared with the
Hawking's coefficient $1/4$. This factor can be absorbed by choosing a
suitable normalization constant when we apply the partition of the integer $%
N $. In any case we are led to the Bekenstein-Hawking relation between
entropy and area\cite{J.Bekenstein,S.Hawking}
\begin{equation}
S=\frac{A}{4l_{p}^{2}}.  \label{p33}
\end{equation}
We can use Eq.$\left( \ref{p33a}\right) $ to compute the entropy for some
specific geometries, for example, the Schwarzschild geometry
\begin{equation}
S=\frac{4\pi \left( 2MG\right) ^{2}}{4G}=4\pi M^{2}G=4\pi
M^{2}l_{p}^{2}=N\ln 2.  \label{p33b}
\end{equation}
Thus the Schwarzschild black hole mass is {\it quantized} in terms of $l_{p}$
giving therefore the relation
\begin{equation}
M=\frac{\sqrt{N}}{2l_{p}}\sqrt{\frac{\ln 2}{\pi }},  \label{p34}
\end{equation}
which is in agreement with the results presented in Refs.\cite%
{Ahluwalia,Hod,Makela,VazWit,Mazur1,Kastrup,JGB}. This implies also that the
level spacing of the transition frequencies is
\begin{equation}
\omega _{0}=\Delta M=\left( 8\pi Ml_{p}^{2}\right) ^{-1}\ln 2
\end{equation}
and the Schwarzschild radius is {\it quantized} in terms of $l_{p}.$ Indeed
\begin{equation}
R_{S}=2MG=2Ml_{p}^{2}=\sqrt{N}l_{p}\sqrt{\frac{\ln 2}{\pi }}.
\end{equation}

\section{The cosmological constant}

\label{p4}Einstein introduced his cosmological constant $\Lambda _{c}$ in an
attempt to generalize his original field equations. The modified field
equations are
\begin{equation}
R_{\mu \nu }-\frac{1}{2}g_{\mu \nu }R+\Lambda _{c}g_{\mu \nu }=8\pi GT_{\mu
\nu }.
\end{equation}%
By redefining
\begin{equation}
T_{tot}^{\mu \nu }\equiv T^{\mu \nu }-\frac{\Lambda _{c}}{8\pi G}g^{\mu \nu
},
\end{equation}%
one can regain the original form of the field equations
\begin{equation}
R_{\mu \nu }-\frac{1}{2}g_{\mu \nu }R=8\pi GT_{\mu \nu },
\end{equation}%
at the prize of introducing a vacuum energy density and vacuum stress-energy
tensor
\begin{equation}
\rho _{\Lambda }=\frac{\Lambda _{c}}{8\pi G};\qquad T_{\Lambda }^{\mu \nu
}=-\rho _{\Lambda }g^{\mu \nu }.
\end{equation}%
If we look at the Hamiltonian in presence of a cosmological term, we have
the expression
\begin{equation}
H=\int_{\Sigma }d^{3}x(N\left( {\cal H}+\rho _{\Lambda }\sqrt{g}\right)
{\cal +}N^{i}{\cal H}_{i}),  \label{p41}
\end{equation}%
where ${\cal H}$ is the usual Hamiltonian density defined without a
cosmological term. We know that the effect of vacuum fluctuation is to
inducing a cosmological term. Indeed by looking at Eq.$\left( \ref{p29}%
\right) $, we have that
\begin{equation}
\frac{\left\langle \Delta H\right\rangle }{V}=-N_{w}\frac{\Lambda ^{4}}{%
64e\pi ^{2}}.  \label{p42}
\end{equation}%
On the other hand, the WDW equation in presence of a cosmological constant
is
\begin{equation}
\left[ \frac{16\pi l_{p}^{2}}{\sqrt{g}}G_{ijkl}\pi ^{ij}\pi ^{kl}-\frac{%
\sqrt{g}}{16\pi l_{p}^{2}}\left( R-2\Lambda _{c}\right) \right] \Psi \left[
g_{ij}\right] =0.
\end{equation}%
By integrating over the hypersurface $\Sigma $ and looking at the
expectation values computed on the state $\left| \Psi _{F}\right\rangle $ of
Eq.$\left( \ref{p30}\right) $ the WDW equation becomes
\[
\left\langle \Psi _{F}\left| \int_{\Sigma }d^{3}x\left[ \frac{16\pi l_{p}^{2}%
}{\sqrt{g}}G_{ijkl}\pi ^{ij}\pi ^{kl}-\frac{\sqrt{g}}{16\pi l_{p}^{2}}R%
\right] \right| \Psi _{F}\right\rangle
\]%
\begin{equation}
=\left\langle \Psi _{F}\left| -\frac{\Lambda _{c}}{8\pi l_{p}^{2}}%
\int_{\Sigma }d^{3}x\sqrt{g}\right| \Psi _{F}\right\rangle =-\frac{\Lambda
_{c}}{8\pi l_{p}^{2}}\int_{\Sigma }d^{3}x\sqrt{g}=-\frac{\Lambda _{c}}{8\pi
l_{p}^{2}}V_{c}.  \label{p43}
\end{equation}%
$V_{c}$ is the cosmological volume. The first term of Eq.$\left( \ref{p43}%
\right) $ is formally the same that generates the vacuum fluctuation $\left( %
\ref{p42}\right) $. Thus, by comparing the second term of Eq.$\left( \ref%
{p43}\right) $ with Eq.$\left( \ref{p42}\right) $, we have
\begin{equation}
-\frac{\Lambda _{c}}{8\pi l_{p}^{2}}V_{c}=-N_{w}\frac{\Lambda ^{4}}{64e\pi
^{2}}V_{w}.  \label{p44}
\end{equation}%
Therefore
\begin{equation}
\Lambda _{c}=N_{w}^{2}\frac{\Lambda ^{4}l_{p}^{2}}{V_{c}8e\pi }V_{w}.
\end{equation}%
The cosmological volume has to be rescaled in terms of the wormhole radius,
in such a way to obtain that $V_{c}\rightarrow N_{w}^{3}V_{w}$  and we have
rescaled the Planck length as in page \pageref{page}. This is the direct
consequence of the boundary rescaling, namely $R_{\pm }$ $\rightarrow $ $%
R_{\pm }/N_{w}$. Thus
\begin{equation}
\Lambda _{c}=\frac{\Lambda ^{4}l_{p}^{2}}{N_{w}8e\pi }.  \label{p45}
\end{equation}%
This is the value of the {\it induced cosmological constant}. On the other
hand, if we apply the area quantization procedure of Eq.$\left( \ref{p33b}%
\right) $ to the de Sitter geometry, one gets
\begin{equation}
S=\frac{3\pi }{l_{p}^{2}\Lambda _{c}}=N\ln 2,
\end{equation}%
that is\footnote{%
A relation relating $\Lambda $ and $G$, via an integer $N$ appeared also in
Ref.\cite{Nojiri}. Nevertheless in Ref.\cite{Nojiri}, $N$ represents the
number of scalar fields and the bound from above and below
\[
\left| 2GN\Lambda /3-2\right| \geq \sqrt{3}
\]%
comes into play, instead of the equality (\ref{p46}).}
\begin{equation}
\frac{3\pi }{\ln 2l_{p}^{2}N}=\Lambda _{c}.  \label{p46}
\end{equation}%
Thus the cosmological constant $\Lambda $ is ``{\it quantized''} in terms of
$l_{p}$. Note that when the wormholes number $N$ is quite ``{\it large}'', $%
\Lambda \rightarrow 0.$ We could try to see what is the rate of change
between an early universe value of the cosmological constant and the value
that we observe. In inflationary models of the early universe is assumed to
have undergone an early phase with a large effective $\Lambda \sim \left(
10^{10}-10^{11}GeV\right) ^{2}$ for GUT era inflation, or $\Lambda \sim
\left( 10^{16}-10^{18}GeV\right) ^{2}$ for Planck era inflation. A
subsequent phase transition would then produce a region of space-time with $%
\Lambda \leq \left( 10^{-42}GeV\right) ^{2}$, i.e. the space in which we now
live. For GUT era inflation, we have (we are looking only at the order of
magnitude)
\begin{equation}
10^{20}-10^{22}GeV^{2}=\frac{1}{N}10^{38}GeV^{2}\rightarrow
N=10^{16}-10^{18},
\end{equation}%
while for Planck era inflation we have
\begin{equation}
10^{32}-10^{36}GeV^{2}=\frac{1}{N}10^{38}GeV^{2}\rightarrow N=10^{6}-10^{2},
\end{equation}%
to be compared with the value of $\left( 10^{-42}GeV\right) ^{2}$ which
gives a wormholes number of the order of
\begin{equation}
10^{-84}GeV^{2}=\frac{1}{N}10^{38}GeV^{2}\rightarrow N=10^{122}.  \label{p47}
\end{equation}%
In our model this very huge number represents the maximum wormholes number
of Planck size that can be stored into an area of radius equal to the
cosmological radius. This is in agreement with observational data of Eq.$%
\left( \ref{i2}\right) $. If we compare the previous value of $\Lambda _{c}$
with the value of Eq.$\left( \ref{p46}\right) $, one gets
\begin{equation}
\Lambda _{c}=\frac{\Lambda ^{4}l_{p}^{2}}{N_{w}8e\pi }=\frac{3\pi }{\ln
2l_{p}^{2}N_{w}},
\end{equation}%
namely we have a constraint on the U.V. cut-off
\begin{equation}
\Lambda ^{4}=\frac{24e\pi ^{2}}{\ln 2l_{p}^{4}}.
\end{equation}%
The probability to realize a foamy spacetime is measured by
\begin{equation}
\Gamma _{{\rm N-holes}}=\frac{P_{{\rm N-holes}}}{P_{{\rm flat}}}\simeq \frac{%
P_{{\rm foam}}}{P_{{\rm flat}}}.
\end{equation}%
In a Euclidean time this is
\begin{equation}
P\sim \left| e^{-\left( \Delta E\right) \left( \Delta t\right) }\right|
^{2}\sim \left| \exp \left( N_{w}\frac{\Lambda ^{4}}{e64\pi ^{2}}\right)
\left( V\Delta t\right) \right| ^{2}.
\end{equation}%
From Eq.(\ref{p44}), we obtain
\begin{equation}
P\sim \left| \exp \left( \frac{\Lambda _{c}}{8\pi l_{p}^{2}}V_{c}\right)
\left( \Delta t\right) \right| ^{2}.
\end{equation}%
To be concrete we can consider again the de Sitter case. Thus
\begin{equation}
\Delta t=2\pi \sqrt{\frac{3}{\Lambda _{c}}}
\end{equation}%
and the cosmological volume is given by
\begin{equation}
V_{c}=\frac{4\pi }{3}\left( \sqrt{\frac{3}{\Lambda _{c}}}\right) ^{3},
\end{equation}%
namely
\begin{equation}
\exp \left( 3\pi /l_{p}^{2}\Lambda _{c}\right) .
\end{equation}%
Thus we recover the Hawking result about the cosmological constant
approaching zero\cite{S.Hawking}. Note that the vanishing of $\Lambda _{c}$
is related to the growing of the wormholes number.

\section{Conclusions}

\label{p5}

In this paper we have continued the investigation of our spacetime foam
model presented in Refs.\cite{Remo1,Remo2}, where we have obtained a ``{\it %
quantization}'' process in the sense that we can fill spacetime with a given
integer number of disjoint non-interacting wormholes. At first look, it
seems that our foam model \ looks promising, since in this framework we have
reproduced certain features that a quantum theory of gravity must possess.
Nevertheless a lot of points must be clarified. First of all the r\^{o}le of
the Planckian cutoff that here is computed by comparing a tree level
quantity (the entropy of \ the de Sitter space) with a one-loop quantity
(the induced cosmological constant or the Casimir energy). Secondly, the
effect of quantum fluctuation has to be inserted in the entropy computation.
This could cause a modification of Eqs.(\ref{p34}) and (\ref{p46}) and
therefore of estimate (\ref{p47}). On the other hand, as a first consequence
we have obtained that the area operator has a discrete spectrum, whose
quanta are Planck size wormholes. This is in agreement with the quantized
area proposed heuristically by Bekenstein and also with the loop quantum
gravity predictions of Refs.\cite{Rovelli,Rovelli1}. Note that in order to
have stability, it is the energy configuration that forces spacetime to be
filled with $N$ wormholes of the Planckian size. Since the area is related
to the entropy via the Bekenstein-Hawking relation, as a direct application,
a ``{\it mass quantization''} of a Schwarzschild black hole whose mass is $M$
is obtained, in agreement with Refs.\cite%
{Ahluwalia,Hod,Makela,VazWit,Mazur1,Kastrup,JGB}. The second consequence of
our model is the generation of a positive cosmological constant induced by
vacuum fluctuations. Due to the uncertainty relation
\begin{equation}
\Delta E\propto \frac{A}{L^{4}}\propto -N_{w}\frac{V}{64\pi ^{2}}\frac{%
\Lambda ^{4}}{e}\propto A\Lambda ^{4}.
\end{equation}%
The negative fourth power of the cutoff (or the inverse of the fourth power
of the region of dimension L) is a clear signal of a Casimir-like energy
generated by vacuum fluctuations. As a consequence a {\it positive
cosmological constant} is induced by such fluctuations.


\begin{references}
\bibitem{Wheeler} J.A. Wheeler, Ann. Phys. {\bf 2 }(1957) 604; J.A. Wheeler,
{\it Geometrodynamics}. Academic Press, New York, 1962.

\bibitem{GAC} G. Amelino-Camelia, Nature {\bf 398} (1999) 216,
gr-qc/9808029; G. Amelino-Camelia, Nature {\bf 393} (1999) 763,
astro-ph/9712103; G. Amelino-Camelia, {\it Gravity-wave interferometers as
probes of a low-energy effective quantum gravity}, gr-qc/9903080.

\bibitem{Yau} R. Schoen, S.T. Yau, Commun. Math. Phys. {\bf 65}, 45 (1979);
Commun. Math. Phys. {\bf 79}, 231 (1981).

\bibitem{GPY} D.J. Gross, M.J. Perry and L.G. Yaffe, Phys. Rev. {\bf D\ 25},
(1982) 330.

\bibitem{HartleHorowitz} J.B. Hartle and G.T. Horowitz, Phys. Rev. D {\bf 24}%
, (1981) 257.

\bibitem{CraneSmolin} L. Crane and L. Smolin, Nucl. Phys. {\bf B} 267 (1986)
714.

\bibitem{Remo} R. Garattini, Int. J. Mod. Phys. {\bf A 18} (1999) 2905,
gr-qc/9805096.

\bibitem{Nariai} S. Nariai, {\it On some static solutions to Einstein's
gravitational field equations in a spherically symmetric case. }Science
Reports of the Tohoku University, {\bf 34} (1950) 160; S. Nariai, {\it On a
new cosmological solution of Einstein's field equations of gravitation. }%
Science Reports of the Tohoku University, {\bf 35} (1951) 62.

\bibitem{Bousso-Hawking} R. Bousso and S.W. Hawking, Phys. Rev. {\bf D} {\bf %
52}, 5659 (1995), gr-qc/9506047; R. Bousso and S.W. Hawking, Phys.Rev. {\bf D%
} {\bf 54} 6312 (1996), gr-qc/9606052.

\bibitem{Modanese} G. Modanese, Phys.Lett. {\bf B }460 (1999) 276,
gr-qc/9906023.

\bibitem{Witten} E. Witten, Nucl. Phys. {\bf B} {\bf 195 (}1982) 481.

\bibitem{Coleman} S. Coleman, Nucl. Phys. {\bf B} {\bf 298} (1988), 178.

\bibitem{Mazur} P.O. Mazur Mod. Phys. Lett. {\bf A} {\bf 4}, (1989) 1497.

\bibitem{BES} R. Brout, F. Englert and P. Spindel, Phys. Rev. Lett. {\bf 43}
(1979) 417.

\bibitem{BES1} R. Brout et al., Nucl. Phys. {\bf B} {\bf 170} (1980), 228.

\bibitem{MVisser} M. Visser, {\it Lorentzian Wormholes} (AIP Press, New
York, 1995) 64.

\bibitem{Remo1} R. Garattini, Phys. Lett. {\bf B 446} (1999) 135,
hep-th/9811187.

\bibitem{Remo2} R. Garattini, Phys. Lett. {\bf B 459} (1999) 461,
hep-th/9906074; R. Garattini, Nucl. Phys. Proc. Suppl. {\bf 88}, 297-300
(2000){\it , }gr-qc/9910037.

\bibitem{BrownYork} J.D. Brown and J.W. York, Phys. Rev. {\bf D} {\bf 47},
1407 (1993).

\bibitem{FroMar} V.P. Frolov and E.A. Martinez, Class.Quant.Grav.{\bf 13}
:481-496,1996, gr-qc/9411001.

\bibitem{HawHor} S. W. Hawking and G. T. Horowitz, Class. Quant. Grav. {\bf %
13 }1487, (1996), gr-qc/9501014.

\bibitem{Martinez} E.A. Martinez, {\it Entropy of eternal black holes}. To
appear in the proceedings of the Sixth Canadian Conference on General
Relativity and Relativistic Astrophysics, gr-qc/9508057.

\bibitem{B.Allen} B. Allen, Phys. Rev. {\bf D} {\bf 30} (1984) 1153.

\bibitem{J.Bekenstein} J. Bekenstein, Phys. Rev. {\bf D7} (1973) 2333.

\bibitem{S.Hawking} S. Hawking, Phys. Lett. {\bf B 134} (1984) 403.

\bibitem{Bekenstein} J. D. Bekenstein, Lett. Nuovo Cimento, {\bf 4} (1972),
737; J. D. Bekenstein, {\it Black Holes: Classical Properties,
Thermodynamics and Heuristic Quantization,} Lectures delivered at the IX
Brazilian School on Cosmology and Gravitation, Rio de Janeiro
7-8/98.gr-qc/9808028.

\bibitem{Mukohyama} S. Mukohyama, {\it The origin of black hole entropy},
gr-qc/9812079.

\bibitem{Hollmann} H. Hollmann, Jou. Math. Phys. {\bf 39}, 11 (1998) 6082,
gr-qc/9610042.

\bibitem{Cavaglià} M. Cavagli\`{a}, V. de Alfaro, A. T. Filippov, Int. J.
Mod. Phys. {\bf D 5} (1996) 227, gr-qc/9508062.M. Cavagli\`{a}, V. de
Alfaro, A. T. Filippov, Int. J. Mod. Phys. {\bf D 4} (1995) 661,
gr-qc/9411070.

\bibitem{Rovelli} C. Rovelli, ``{\it Loop Quantum Gravity''}, gr-qc/9710008.

\bibitem{Rovelli1} C. Rovelli, Phys. Rev. Lett. {\bf 77} (1996) 3288,
gr-qc/9603063;

\bibitem{Ahluwalia} D.V. Ahluwalia, Int.J.Mod.Phys. D {\bf 8} (1999) 651,
astro-ph/9909192.

\bibitem{Hod} S. Hod, Phys. Rev. Lett. {\bf 81}, 4293 (1998), gr-qc/9812002.

\bibitem{Makela} J. M\"{a}kel\"{a}, {\it Schr\"{o}dinger Equation of the
Schwarzschild Black Hole}, gr-qc/9602008; J. M\"{a}kel\"{a}, Phys.Lett. {\bf %
B390} (1997) 115-118.

\bibitem{VazWit} C. Vaz and L. Witten, Phys.Rev. {\bf D 60} (1999) 024009,
gr-qc/9811062.

\bibitem{Mazur1} P.O. Mazur, Acta Phys.Polon. {\bf 27} (1996) 1849-1858,
hep-th/9603014.

\bibitem{Kastrup} H.A. Kastrup, Phys.Lett. {\bf B 413} (1997) 267-273,
gr-qc/9707009; H.A. Kastrup, Phys.Lett. {\bf B 419} (1998) 40-48,
gr-qc/9710032.

\bibitem{JGB} J. Garcia-Bellido, {\it QUANTUM BLACK HOLES}, hep-th/9302127.

\bibitem{Nojiri} S. Nojiri, O. Obregon, S.D. Odintsov and K.E. Osetrin,
Phys.Lett. {\bf B 449} (1999) 173-179, hep-th/9812164.
\end{references}
\end{document}